\documentclass[twocolumn,epjc3]{svjour3}  

\usepackage{lineno}

\usepackage{graphicx}

\begin{document}

\title{Large area Si low-temperature light detectors with Neganov-Luke effect}

\author{
{M.~Biassoni}\thanksref{UNIMIB,INFN-MIB}
\and
{C.~Brofferio}\thanksref{UNIMIB,INFN-MIB}
\and
{S.~Capelli}\thanksref{UNIMIB,INFN-MIB}
\and
{L.~Cassina}\thanksref{UNIMIB,INFN-MIB}
\and
{M.~Clemenza}\thanksref{UNIMIB,INFN-MIB}
\and
{O.~Cremonesi}\thanksref{INFN-MIB}
\and
{M.~Faverzani}\thanksref{UNIMIB,INFN-MIB}
\and
{E.~Ferri}\thanksref{UNIMIB,INFN-MIB}
\and
{A.~Giachero}\thanksref{INFN-MIB}
\and
{L.~Gironi}\thanksref{UNIMIB,INFN-MIB,e1}
\and
{C.~Giordano}\thanksref{FBK,UNIMIB}
\and
{C.~Gotti}\thanksref{UNIMIB,INFN-MIB}
\and
{M.~Maino}\thanksref{UNIMIB,INFN-MIB}
\and
{B.~Margesin}\thanksref{FBK}
\and
{A.~Nucciotti}\thanksref{UNIMIB,INFN-MIB}
\and
{M.~Pavan}\thanksref{UNIMIB,INFN-MIB}
\and
{G.~Pessina}\thanksref{INFN-MIB}
\and
{E.~Previtali}\thanksref{INFN-MIB}
\and
{A.~Puiu}\thanksref{INFN-MIB}
\and
{M.~Sisti}\thanksref{UNIMIB,INFN-MIB}
\and
{F.~Terranova}\thanksref{UNIMIB,INFN-MIB}
}

\thankstext{e1}{e-mail: luca.gironi@mib.infn.it}

\institute{{Dipartimento di Fisica - Universit\`{a} di Milano-Bicocca, Milano, Italy}\label{UNIMIB}
\and
{INFN - Sezione di Milano-Bicocca, Milano, Italy}\label{INFN-MIB}
\and
{Fondazione Bruno Kessler, Trento, Italy }\label{FBK}
}

\maketitle

\abstract{Next generation calorimetric experiments for the search of rare events rely on the detection of tiny amounts of light (of the order of 20 optical photons) to discriminate and reduce background sources and improve sensitivity. Calorimetric detectors are the simplest solution for photon detection at cryogenic (mK) temperatures. The development of silicon based light detectors with enhanced performance thanks to the use of the Neganov-Luke effect is described. The aim of this research line is the production of high performance detectors with industrial-grade reproducibility and reliability.}



\section{Introduction}

In the last decade low temperature thermal detectors have proved to be powerful tools for high sensitivity spectroscopy in particle and nuclear physics. Calorimeters are used in many of the most sensitive experiments for the search for rare events, like double beta decay, rare nuclear decays and dark matter direct detection. On the path to the improvement of their sensitivity, the reduction of the background is one of the strategies to be pursued, and hybrid detectors for the active rejection of spurious events are the most promising option. The simultaneous detection of the heat and light produced by an interaction in the absorbing crystal makes particle identification possible, both in scintillating calorimeters \cite{scintillanti,cresst} (where the light yields of nuclear recoils, alpha and beta/gamma particles are different) and in non-scintillating calorimeters by means of ancillary scintillating structures \cite{absurd} or by the detection of Cherenkov photons \cite{cherenkov}. In all these cases, especially when the corresponding technology has to be applied to future generation experiments, high detection efficiency and low threshold are required as the amount of light to be measured is as small as $\sim$50 eV. At the same time scalability and reproducibility are mandatory when dealing with ton-scale experiments with hundreds to thousands of channels.

\section{Calorimetric Light Detectors}

In order to detect the photons emitted by large mass crystals operated as calorimeters at few mK, other calorimeters are usually the most effective tool.  Presently running experiments or advanced R\&Ds \cite{cresst,lucifer} use calorimetric light detectors of various materials (germanium, SOS - silicon on sapphire) and temperature reading techniques (Ge NTD thermistors, TES, composite TES). These detectors proved good to excellent performance in terms of energy resolution and thresholds down to hundreds of eV. 
Even though germanium slabs are commercial products, the semiconductor industry is almost completely focused on silicon substrates. Silicon on sapphire detectors are also not very common outside the scientific environment. As a consequence, silicon is much cheaper and the available technological expertise of the semiconductor industry is wider. Silicon detectors with Neganov-Luke enhanced thermal signal can therefore be produced with reproducible processes and their properties tuned with high precision. Silicon has also a wider range of processing technologies, including micromachining processes, that potentially allow the integration of thermal sensors and mechanical suspension structures.
A further advantage of silicon over germanium is the fact that specific heat of silicon is a factor $\sim$4.5 smaller than germanium \cite{Si_SpecificHeat}, opening the possibility of building substantially larger detectors without compromising the signal amplitude which is inversely proportional to the heat capacity of the device.

\subsection{Neganov-Luke amplification}

The Neganov-Luke effect \cite{naganov-luke} is a mechanism that leads to thermal signal amplification in semiconductor calorimeters when a static electric field is applied. In standard operating conditions, the signal is the temperature rise generated by the conversion of the energy deposited by an interacting particle into heat. Under the assumption that all the energy deposited by the particle converts into thermal phonons in a time lapse much smaller than the typical time response of the temperature sensors, the amplitude of the temperature variation is directly proportional to the originally deposited energy. Given the large number of information carriers (thermal phonons) in the calorimeter, the intrinsic energy resolution is beyond reach of any other detector. The energy resolution that can realistically be achieved is limited by imperfect thermalisation mechanisms and temperature sensor performance, but it is still of the order of 100 eV for small size light detectors.

When a static electric field is applied to the semiconductor absorber of the calorimeter, the electron-hole pairs produced by the primary particle interactions drift and acquire kinetic energy. This energy is converted into heat while the pairs scatter the absorber lattice during their motion, producing an additional temperature increase, or thermal signal gain.
As long as the electron-hole pairs drift and are collected without significant recombination, the additional heat produced is proportional to the number of pairs and to the voltage applied to generate the electric field. The total energy resulting in a thermal signal is therefore:

\begin{eqnarray}\label{eq:luke gain}
E_{\mathrm{tot}} & = & E - E_{\mathrm{production}}+ E_{\mathrm{field}}\nonumber \\ 
& = & E - n_{\mathrm{e-H}}\,\ \delta + n_{\mathrm{e-H}}\,\, q V \nonumber \\
& = & E + \frac{E}{\epsilon} \,\, (-\delta + q V)\nonumber \\
& = & E\left(1 -\frac{\delta}{\epsilon} + \frac{qV}{\epsilon} \right) \mathrm.
\end{eqnarray}

\bigskip

where $E$ is the energy deposited by the particle interaction, $E_{\mathrm{production}}$ is the fraction of energy used to produce electron-hole pairs, while $E_{\mathrm{field}}$ is the energy of the electric field converted into heat to enhance the thermal signal. The former equals the number of electrons and holes ($n_{\mathrm{e-H}}$) times the band gap $\delta$ (1.17 eV for silicon \cite{Si_BandGap}), while the latter is $n_{\mathrm{e-H}}$ times the charge of the electron ($q$) multiplied by the potential $V$. The number of electron-hole pairs equals the energy originally deposited divided by the mean energy $\epsilon$ needed to produce a pair in the considered semiconductor. In the case of silicon this energy is $\sim$3.8 eV at 77 K.

It is interesting to note that, as soon as \textit{V} is larger than a few volts, the last term of Equation~\ref{eq:luke gain} quickly becomes dominant. In this regime the thermal detector is working like a ionization detector because the thermal signal is almost entirely generated by the drifting charges, and its energy resolution is therefore expected to be determined by the fluctuations in the number of electron-hole pairs generated in the primary interaction. As far as the noise is not affected and the pulse amplitude is increased, the detection threshold is reduced by the same factor.

Some literature exists where attempts to exploit this effect to amplify the signal produced by light absorbed in calorimetric detectors are described \cite{Stark,Isaila12}. In this work we report on similar measurements performed with different electrode layouts, in a wider voltage range and with carefully selected high resistivity silicon substrates.

\begin{figure*}[t]
\begin{minipage}[t]{0.5\textwidth}
 \includegraphics[height=55mm]{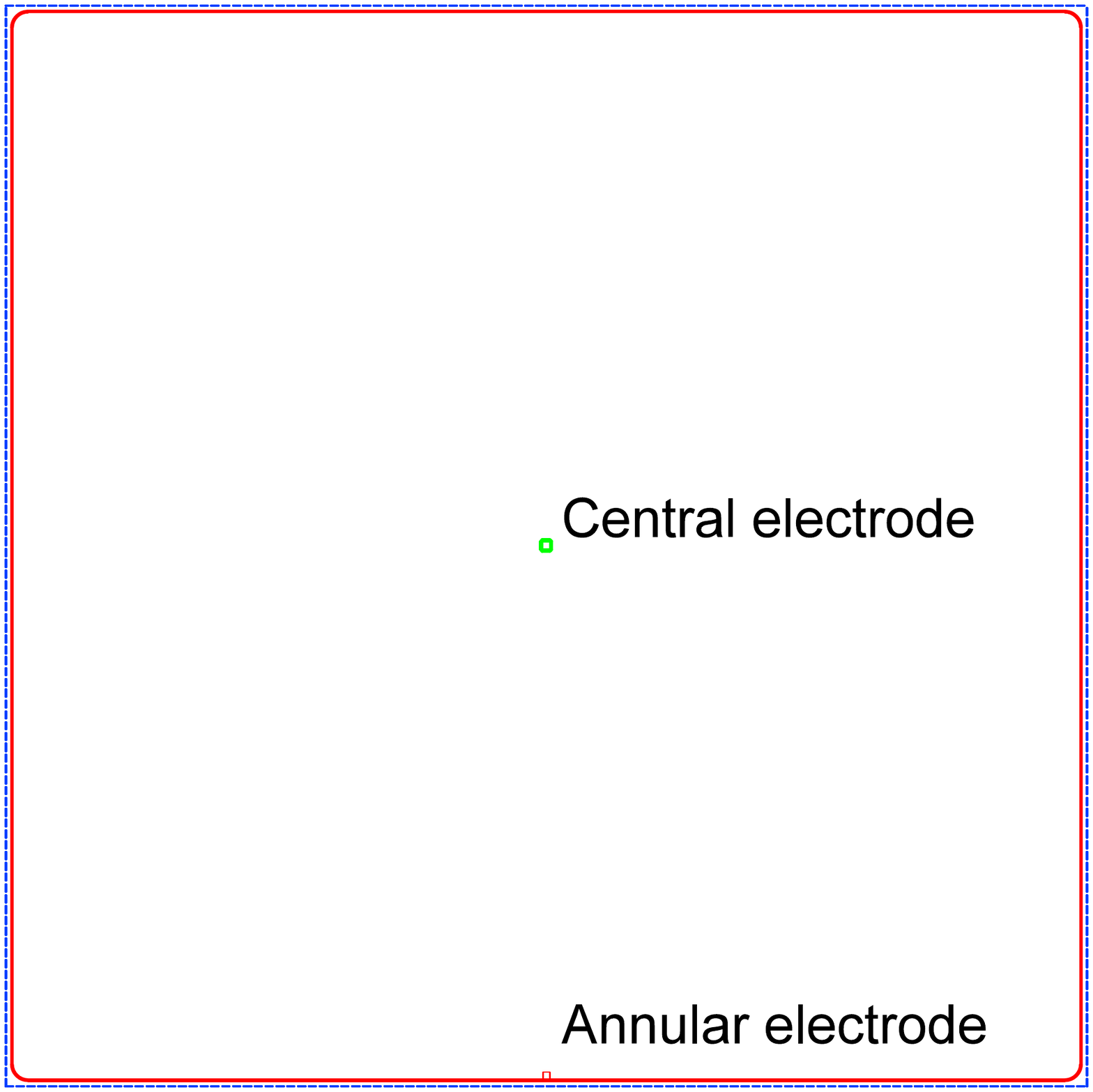}
 \end{minipage}
 \begin{minipage}[t]{0.5\textwidth}
 \includegraphics[height=55mm]{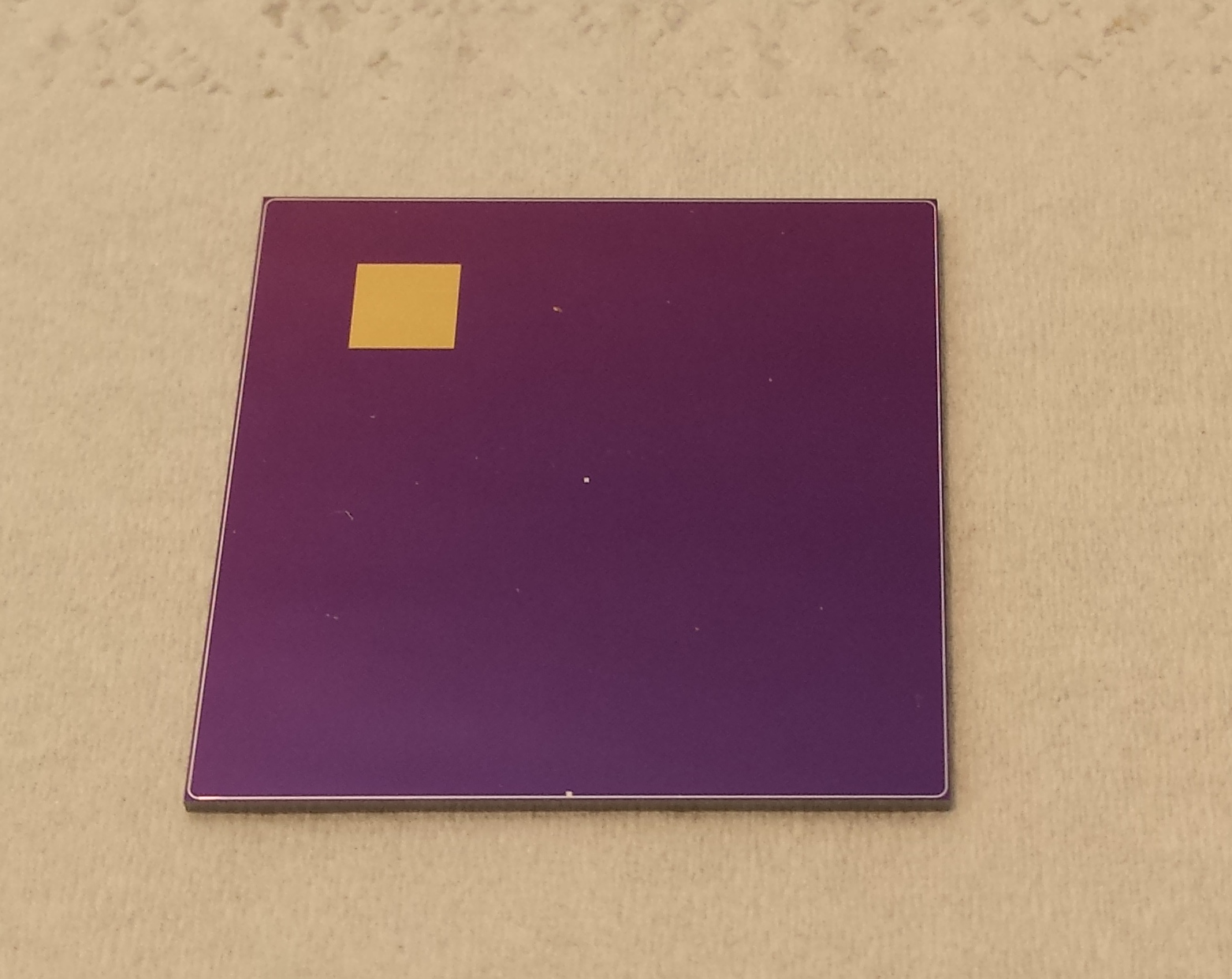}
 \end{minipage}
\caption{Neganov-Luke detector geometry. Left: technical drawing of the electrode layout. Right: picture of the actual detector with aluminum electrodes and bonding pads. The central dot is the bonding pad and the electrode itself. The annular electrode has a pad for wire bonding at the midpoint of one side. The square in the upper left corner is a gold patch foreseen for the placement of the NTD temperature sensor.}
\label{fig:LukeGeom}
\end{figure*}

\section{Experimental setup}

The detectors we discuss are Si crystals held by means of PTFE pieces fixed to a copper frame. Each of them consists of a 20x20 mm$^2$, 625 $\mu$m thick, crystalline Si absorber. The room-temperature resistivity of the Si is $>5$ k$\Omega \cdot$cm. For the application of the Neganov-Luke voltage the absorber is equipped with two Al electrodes directly evaporated onto its surface. Below the electrodes the silicon substrate is heavily doped in order to build a ohmic contact. We tested different geometries and the best results were obtained with the geometry shown in Figure~\ref{fig:LukeGeom}.

The detectors were installed in an Oxford Instrument TL200 dilution refrigerator located in the Cryogenic Laboratory of Milano-Bicocca University. 

The temperature sensor is a 3x1x0.5 mm$^3$ Neutron Transmutation Doped (NTD) germanium thermistor. It is thermally coupled to the Si crystal via glue spots of Araldite. When the NTDs are biased with a polarizing current, the strong dependence of their resistance on the temperature translates in a variation of the voltage across the thermistor each time a temperature pulse is recorded. This electrical signal (voltage) is amplified, filtered to remove high frequency noise, digitized and continuously recorded.
The off-line analysis allows to determine the pulse amplitude, as well as many pulse shape parameters.

Two types of light sources have been faced to the detectors: BGO (Bismuth Germanate) crystals illuminated by a $^{232}$Th source located outside the cryostat and YAP (Yttrium Aluminum Perovskite) scintillators with a monochromatic $\alpha$ source deposited on a surface.

\section{Detector performance}

The performance of the detectors have been studied by applying different polarizing voltages at the Neganov-Luke effect electrodes. The polarity of the bias has also been inverted to characterize any asymmetric behaviour of the device and determine the best operating conditions. Voltages up to 240 V have been applied on 12 detectors without encountering any failure and observing a consistent behaviour.

\subsection{Detectors noise and energy resolution}
The noise level of the detectors has been evaluated during the data taking at different values of Neganov-Luke voltage by randomly sampling the baseline after the application of an optimal filtering algorithm. No dependence of the noise level on the applied voltage has been observed up to the maximum applied voltage of 240 V. The energy resolution has also been estimated by measuring the width of the peak corresponding to the light pulses produced by the YAP source. As expected, already at the lowest value of applied voltage (40 V) the relative energy resolution is compatible with the width of the light source emission of about 7\% (dominated by the alpha source smearing), and remains constant when the voltage is increased.

\subsection{Pulse amplitude and gain}

\begin{figure*}[t]
\centering
\includegraphics[width=.8\textwidth]{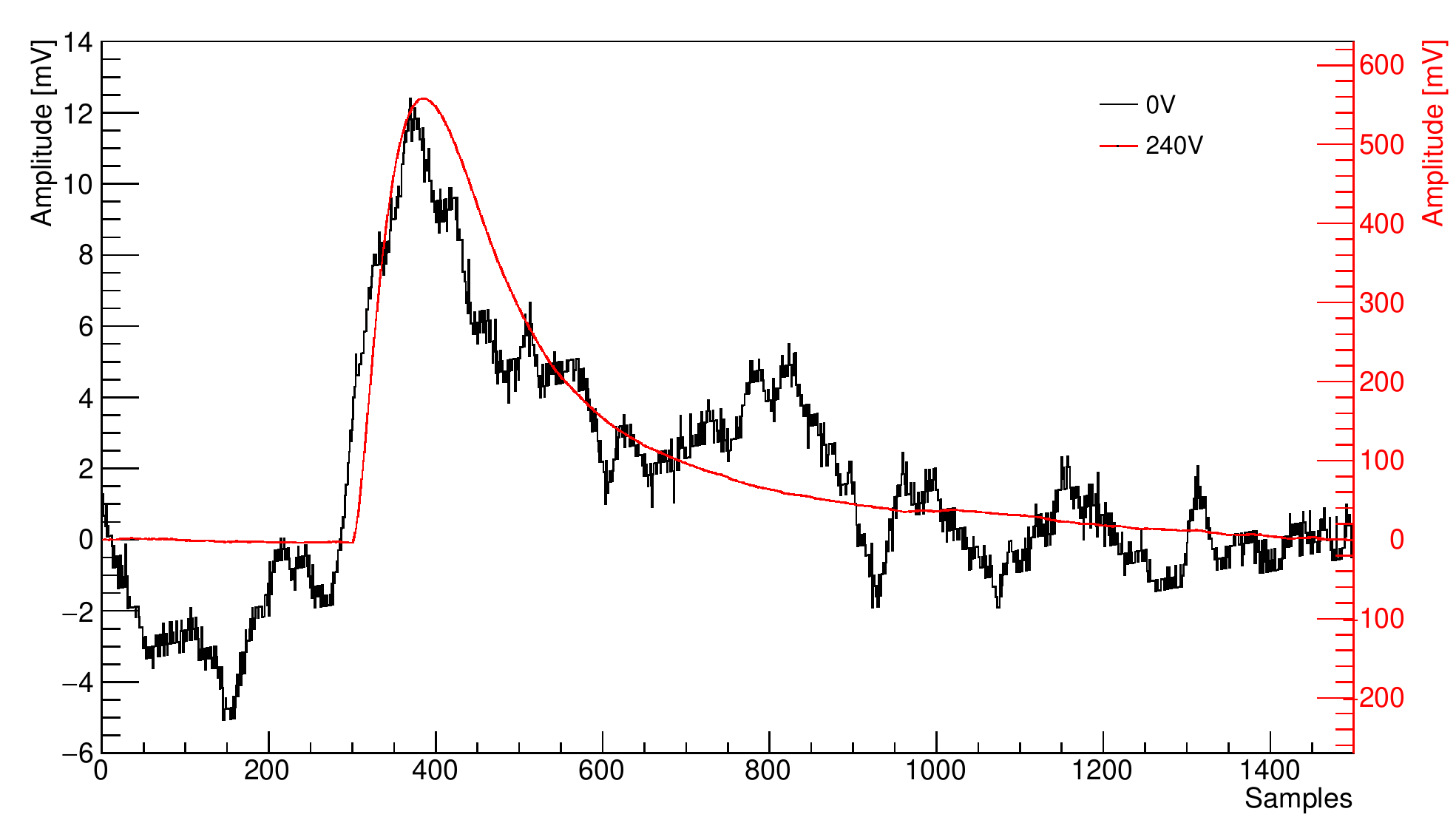}
\caption{Comparison of pulses with and without Neganov-Luke amplification: black line represents a pulse without any amplification while the red line (and red scale on the right) is a pulse corresponding to the same energy deposition amplified by a 240 V bias. The signal-to-noise ratio is improved by a factor $\sim60$.}
\label{fig:LukeGain}
\end{figure*}

When a voltage is applied to the electrodes the amplitude of the pulses generated by light interacting with the absorber is largely increased. The corresponding gain $G(V_{\mathrm{Luke}})$, defined as the ratio between the amplitude of a pulse with a given bias voltage $V_{\mathrm{Luke}}$ and the amplitude of a pulse produced by a particle of the same energy with no voltage applied, shows a smooth and nearly linear dependence on the voltage, with absolute value depending on the electrodes design. The gain stability over time has been characterized and a general behaviour has been identified on all the different electrode designs: after an initial transient during which the gain decreases exponentially, the pulse amplitude stabilizes. The time constant of the transient strongly depends on the rate of energy deposition on the absorber, and hence on the rate of charge produced and drifted by the electric field: the light detectors facing the BGO source with low rate of interaction are stable over long time scales, while the gain of the detectors facing the YAP sources quickly decays by a factor 20-40\%. This behaviour, already observed in similar applications \cite{Stark,Isaila12,CDMS}, is ascribed to charge trapping phenomena in the detector and the consequent generation of a shielding field whose features can be observed and studied when the polarizing voltage is removed. This effect can be reduced with a careful selection of high purity silicon absorbers and by following a proper procedure when the voltage is applied; the residual charge can be easily removed by flashing the detector with light. Given the rate of interaction that the detectors must sustain in rare events experiments (typically few mHz), the gain reduction can be considered negligible.

By applying a Neganov-Luke voltage of 240 V to the detectors described and shown in Figure~\ref{fig:LukeGeom}, a gain of $\sim60$ has been obtained on the pulse amplitude, with no visible effects on the noise level. An equal improvement of the signal-to-noise ratio has therefore been achieved, as shown for example in Figure~\ref{fig:LukeGain}.
Since no misbehaviour has been observed in any of the detectors up to 240 V, higher voltages will be tested in a future measurement campaign with a dedicated cabling to overcome the limits of a standard cryostat wiring system.

\section{Conclusions}

A technique that exploits the Neganov-Luke effect to enhance the thermal signal in cryogenic detectors has been successfully applied to silicon light detectors with a novel design. A large gain in the signal-to-noise ratio has been demonstrated and the good behaviour of the detectors tested in this work proves that further improvements are realistically achievable. Thanks to the very low energy threshold, large detection surface, low cost and performance reproducibility, these detectors are suitable for many applications in the physics of rare events with low temperature detectors, where small amounts of light (scintillation or Cherenkov photons) must be detected in next generation experiments in order to reject the spurious background.

\end{document}